\documentclass[final,5p,times,twocolumn]{elsarticle}

\usepackage{graphicx}
\usepackage{amssymb}
\usepackage{amsmath,amsthm}
\usepackage[amssymb]{SIunits}

\begin{document}

\begin{frontmatter}

\title{Effective material parameter retrieval for thin sheets: theory and application to graphene, thin silver films, and single-layer metamaterials}

\author[Iowa]{Philippe Tassin\corref{cor1}}
\ead{tassin@ameslab.gov}
\author[Iowa]{Thomas Koschny}
\ead{koschny@ameslab.gov}
\author[Iowa,Greece]{Costas M. Soukoulis}
\ead{soukoulis@ameslab.gov}

\address[Iowa]{Ames Laboratory---U.S. DOE and Department of Physics and Astronomy, Iowa State University, Ames, IA 50011, USA}
\address[Greece]{Institute of Electronic Structure and Lasers (IESL), FORTH, 71110 Heraklion, Crete, Greece}

\cortext[cor1]{Corresponding author.}

\begin{abstract}
An important tool in the field of metamaterials is the extraction of effective material parameters from simulated or measured scattering parameters of a sample. Here we discuss a retrieval method for thin-film structures that can be approximated by a two-dimensional scattering sheet. We determine the effective sheet conductivity from the scattering parameters and we point out the importance of the magnetic sheet current to avoid an overdetermined inversion problem. Subsequently, we present two applications of the sheet retrieval method. First, we determine the effective sheet  conductivity of thin silver films and we compare the resulting conductivities with the sheet conductivity of graphene. Second, we apply the method to a cut-wire metamaterial with an electric dipole resonance. The method is valid for thin-film structures such as two-dimensional metamaterials and frequency-selective surfaces and can be easily generalized for anisotropic or chiral media.
\end{abstract}


\end{frontmatter}

\section{Introduction}
\label{Sec:Introduction}
The development of metamaterials, i.e., artificial microstructured materials in which small, subwavelength electric circuits replace atoms as the basic unit of interaction with electromagnetic radiation, has been a vibrant research topic in the field of nanophotonics during the past decade~\cite{Smith-2004,Litchinitser-2008,Soukoulis-2010}. Structuring metamaterials on a subwavelength scale makes it possible to create electromagnetic media with properties not found in natural materials, while still allowing to describe them as effectively continuous media with constitutive parameters such as the electric permittivity and the magnetic permeability~\cite{Koschny-2005}. With appropriately designed constituents, it was shown feasible to design metamaterials with exotic material response, e.g., magnetism at terahertz and optical frequencies, simultaneous negative permittivity and negative permeability (the so-called left-handed materials)~\cite{Smith-2000}, giant chirality~\cite{Plum-2009}, and slow-light media~\cite{Papasimakis-2008,Tassin-2009,Liu-2009}. The first metamaterial was fabricated in 2001 by combining metal wires exhibiting negative permittivity and split-ring resonators (SRR) exhibiting negative permeability in a single material~\cite{Shelby-2001}. SRRs are still commonly used in the microwave band, but have been replaced by other magnetically resonant structures such as slab-wire pairs and fishnets at terahertz frequencies and above, since these meta-atoms ease the problem of saturation of the magnetic response at those frequencies~\cite{Zhou-2005} and also simplify the fabrication.

Metamaterials exhibit novel electromagnetic phenomena, such as backward wave propagation, negative refraction, and inverse Doppler effect~\cite{Veselago-1968}. Furthermore, they enable the compensation of the propagation phase and the restoration of the amplitude of evanescent waves in optical structures, resulting in lenses with subwavelength resolution~\cite{Pendry-2000} and photonic devices going beyond the diffraction limit~\cite{Engheta-2002,Kockaert-2006,Alu-2007,Tassin-2008}. Through the technique of transformation optics, metamaterials with arbitrary values of the permittivity and permeability ultimately allow for maximal control over light propagation, which has led to surprising new physics such as invisibility cloaks, beam transformation, and frequency conversion in linear media~\cite{Leonhardt-2006,Pendry-2006,Rahm-2008,Ginis-2010}.

A basic tool in the study of metamaterials is the so-called retrieval method, i.e., the extraction of effective medium parameters corresponding to a metamaterial with given microscopic structure. Effective material parameters are important because they make the link between the microscopic response of metamaterials and the homogeneous macroscopic media assumed in many proposed applications. We assume here that the meta-atoms are sufficiently subwavelength, so that the effective response is only weakly spatially dispersive and, hence, can be described by frequency-dependent polarization and magnetization fields~\cite{Menzel-2010}. For ease of presentation, we assume here (two-dimensional) isotropy in the plane of the thin-film structure, but the considerations below can be straightforwardly generalized for anisotropic samples. 

There are several methods to determine the effective material parameters. One approach involves the averaging of the electromagnetic fields following their definition~\cite{Jackson-1962}; the volume averaging can sometimes be replaced by line and surface averages~\cite{Smith-2006}. These averaging methods and other methods requiring the knowledge of the microscopic fields~\cite{Popa-2005} are, however, often difficult to apply, since it is complicated or even impossible to obtain the microscopic fields. Another approach involves comparing the metamaterial sample with a slab of a homogeneous medium and determining the constitutive parameters of the slab that gives the same scattering matrix as the metamaterial~\cite{Smith-2002,Chen-2004,Smith-2005}. This is attractive, because scattering experiments can be easily set up, both experimentally and numerically. One problem that arises when using the latter retrieval method with thin-film samples is choosing a suitable length of the slab~\cite{Holloway-2009}. Since the thickness of such thin-film samples is typically much smaller than the wavelength, it makes actually more sense to compare it to a two-dimensional current sheet and characterize the sample by sheet susceptibilities~\cite{Holloway-2009} or sheet currents (this paper) that result in the same scattering parameters as obtained for the metamaterial.

\begin{figure}
\begin{center}
\includegraphics[clip]{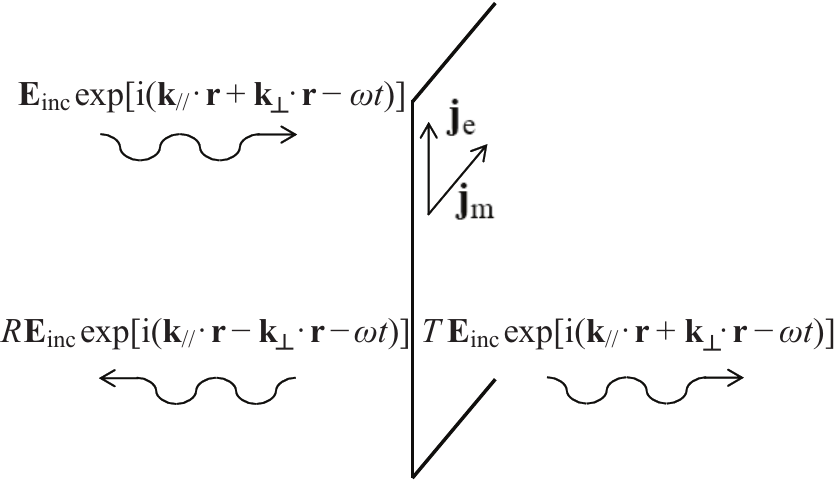}
\end{center}
\caption{Scattering of a plane wave by a two-dimensional current sheet with electric sheet current $\mathbf{j}_\mathrm{e}$ and magnetic sheet current $\mathbf{j}_\mathrm{m}$.}
\label{Fig:CurrentSheet}
\end{figure}

\section{Retrieval method}
\label{Sec:Retrieval}
The two-dimensional retrieval method starts from a rather standard problem of electrodynamics---the scattered fields of a two-dimensional sheet carrying electric and magnetic sheet currents. The problem is sketched in Fig.~\ref{Fig:CurrentSheet}. A plane wave $\mathbf{E}_\mathrm{inc} \exp[\mathrm{i}(\mathbf{k}_{//}\cdot\mathbf{r}+\mathbf{k}_{\perp}\cdot\mathbf{r}-\omega t)]$ illuminates the current sheet and is scattered into a reflected wave $R\:\mathbf{E}_\mathrm{inc} \exp[\mathrm{i}(\mathbf{k}_{//}\cdot\mathbf{r}-\mathbf{k}_{\perp}\cdot\mathbf{r}-\omega t)]$ and a transmitted wave $T\:\mathbf{E}_\mathrm{inc} \exp[\mathrm{i}(\mathbf{k}_{//}\cdot\mathbf{r}+\mathbf{k}_{\perp}\cdot\mathbf{r}-\omega t)]$. Since the current sheet is assumed to have a linear response, we can work in the harmonic regime, and we will drop the factor $\exp(-\mathrm{i}\omega t)$ from now on.

The incident and scattered waves must satisfy the boundary conditions
\begin{align}
\mathbf{n} \cdot \left( \mathbf{D} - \mathbf{D}'\right) &= \rho_\mathrm{e},\label{Eq:BC1}\\
\mathbf{n} \cdot \left( \mathbf{B} - \mathbf{B}'\right) &= \rho_\mathrm{m},\label{Eq:BC2}\\
\mathbf{n} \times \left( \mathbf{E} - \mathbf{E}'\right) &= -\mathbf{j}_\mathrm{m},\label{Eq:BC3}\\
\mathbf{n} \times \left( \mathbf{H} - \mathbf{H}'\right) &= \mathbf{j}_\mathrm{e},\label{Eq:BC4}
\end{align}
where $\mathbf{n}$ is the surface normal of the current sheet, $\mathbf{j}_\mathrm{e}$ and $\mathbf{j}_\mathrm{m}$ are the electric and magnetic sheet currents, respectively, $\mathbf{D}$ is the electric displacement field, $\mathbf{B}$ is the magnetic induction field, $\mathbf{E}$ is the electric field, and $\mathbf{H}$ is the magnetic field. It might seem strange that we include a magnetic sheet current here, but we will see that this is essential to the retrieval method and we will explain below how an effective magnetic sheet current originates in a thin-film sample.

Substituting the incident and scattered fields in Eqs.~(\ref{Eq:BC1})-(\ref{Eq:BC4}), and using the identity $\mathbf{H} = \mathbf{k}\times\mathbf{E}/(\omega\mu_0)$, yields
\begin{align}
\mathbf{j}_\mathrm{e} &= \frac{1}{\omega\mu_0} \left( 1 - R - T \right) \mathbf{n}\times\left(\mathbf{k}\times\mathbf{E}_\mathrm{inc}\right)\nonumber\\
                      &= \zeta^{-1} \left( 1 - R - T \right) \left(\mathbf{E}_\mathrm{inc}\right)_{//},\label{Eq:SheetCurrentE}\\
\mathbf{j}_\mathrm{m} &= -\left( 1 + R - T \right) \mathbf{n}\times\mathbf{E}_\mathrm{inc}\nonumber\\
                      &= \zeta \left( 1 + R - T \right)\left(\mathbf{H}_\mathrm{inc}\right)_{//},\label{Eq:SheetCurrentM}
\end{align}
where $\zeta$ is the wave impedance defined by $\zeta^{-1} = k_\perp / (\omega\mu_0)$, $\eta_0$~is the characteristic impedance of vacuum, and $k_0 = \omega/c$ is the free-space wavenumber.

To find the conductivities, we have to relate the sheet currents to the local fields, $\mathbf{j}_\mathrm{e} \equiv \sigma_{//}^{(\mathrm{e})} \mathbf{E}_\mathrm{loc}$ and $\mathbf{j}_\mathrm{m} \equiv \sigma_{//}^{(\mathrm{m})} \mathbf{H}_\mathrm{loc}$. Since the current sheet carries electric as well as magnetic sheet currents, both the electric and magnetic fields have discontinuities at the current sheet. Therefore, the local fields must be defined as an average across the current sheet:
\begin{align}
\mathbf{E}_\mathrm{loc} &= \frac{\mathbf{E}_{//} + \mathbf{E}_{//}'}{2} = \frac{1}{2} \left( 1 + R + T \right) \left(\mathbf{E}_\mathrm{inc}\right)_{//},\label{Eq:LocalFieldE}\\
\mathbf{H}_\mathrm{loc} &= \frac{\mathbf{H}_{//} + \mathbf{H}_{//}'}{2} = \frac{1}{2} \left( 1 - R + T \right) \left(\mathbf{H}_\mathrm{inc}\right)_{//}.\label{Eq:LocalFieldH}
\end{align}
Combining Eqs.~(\ref{Eq:SheetCurrentE})-(\ref{Eq:SheetCurrentM}) with Eqs.~(\ref{Eq:LocalFieldE})-(\ref{Eq:LocalFieldH}), we arrive at the formulae that give us the effective conductivities that are commensurate with given (measured or simulated) scattering coeffcients:
\begin{align}
\sigma_{//}^{(\mathrm{e})} &= \frac{2}{\zeta} \left( \frac{1-R-T}{1+R+T} \right),\label{Eq:RetrievalElectric}\\
\sigma_{//}^{(\mathrm{m})} &= 2\zeta \left( \frac{1+R-T}{1-R+T} \right).\label{Eq:RetrievalMagnetic}
\end{align}
This is the central result of this paper.

It is interesting to note that these formulae can be uniquely inversed,
\begin{align}
T &= \frac{4-\sigma_{//}^{(\mathrm{e})}\sigma_{//}^{(\mathrm{m})}}{4+2\zeta\sigma_{//}^{(\mathrm{e})} + 2\zeta^{-1}\sigma_{//}^{(\mathrm{m})}+\sigma_{//}^{(\mathrm{e})}\sigma_{//}^{(\mathrm{m})}},\\
R &= -\frac{2\left(\zeta\sigma_{//}^{(\mathrm{e})}-\zeta^{-1}\sigma_{//}^{(\mathrm{m})}\right)}{4+2\zeta\sigma_{//}^{(\mathrm{e})} + 2\zeta^{-1}\sigma_{//}^{(\mathrm{m})}+\sigma_{//}^{(\mathrm{e})}\sigma_{//}^{(\mathrm{m})}},
\end{align}
from which it is obvious that there is an isomorphism between the scattering parameters and the sheet conductivities. Consequently, the thin sheet retrieval does not suffer from the ambiguities encountered in the traditional retrieval method, such as branch cuts, residue classes, or free parameters (the slab thickness). We can now also understand why we need the magnetic sheet current; without the magnetic sheet current, the retrieval problem would in general be overdetermined with only one complex parameter (the electric sheet conductivity) determined by two complex scattering coefficients ($R$ and $T$). Of course, if the sample can really be considered as an electric current sheet, the reflection and transmission coefficients are related by $1+R=T$ and the inversion would not be overdetermined. Nevertheless, it may probably be better in practice to use Eqs.~(\ref{Eq:RetrievalElectric})-(\ref{Eq:RetrievalMagnetic}) and then check whether the magnetic sheet conductance is negligible.

\section{Application to thin silver films and comparison with graphene}
\label{Sec:SilverFilms}
Thin conducting films are frequently employed in nanophotonics. Gold and silver are normally used because of their moderate losses and relatively easy chemical handling. Nevertheless, dissipative losses in those metals are still appreciable at optical frequencies and it is, therefore, interesting to search for alternative conducting materials. Here we want to compare a silver film with graphene. The conducting properties of silver are characterized by its frequency-dependent bulk conductivity in\unit{}{\siemens\per\meter}. As a two-dimensional electronic system, graphene is characterized by its sheet conductivity, which is almost flat and equal to $\sigma_\mathrm{g} = \unit{6.08 \times 10^{-5}}{\siemens}$ starting in the infrared and extending into the visible frequency band.

\begin{figure}
\begin{center}
\includegraphics[clip]{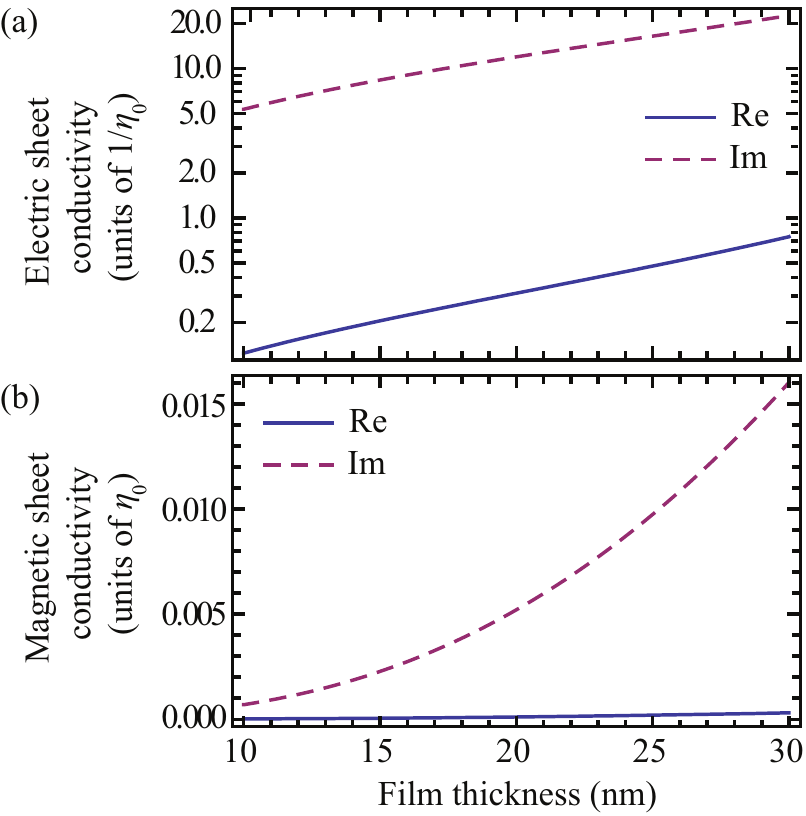}
\end{center}
\caption{(a)~Effective electric sheet conductivity of thin silver films as a function of film thickness at $\lambda = \unit{1.55}{\micro\meter}$. (b)~Effective magnetic sheet conductivity of thin silver films as a function of film thickness at $\lambda = \unit{1.55}{\micro\meter}$.}
\label{Fig:SheetCondSilverFilms}
\end{figure}

In order to compare the current transport in thin silver films and graphene, we use our sheet retrieval method to describe the properties of silver films with thickness between \unit{10}{\nano\meter} and \unit{40}{\nano\meter} by sheet conductivities. We first calculate the scattering parameters for silver films at a wavelength of \unit{1.55}{\micro\meter} using the Drude model from Ordal \textit{et~al.}~\cite{Ordal-1985} for the bulk conductivity (note that the Drude model in Ordal \textit{et~al.}\ has been fitted against experimental data from thin films consistent with the film thickness here) and then we apply Eqs.~(\ref{Eq:RetrievalElectric})-(\ref{Eq:RetrievalMagnetic}). The resulting sheet conductivities are plotted in Fig.~\ref{Fig:SheetCondSilverFilms}(a)-(b). The magnetic sheet conductivity is negligible for the thinnest films, but its imaginary part reaches 1\% of the vacuum impedance at a thickness of approximately \unit{25}{\nano\meter}. For thicker films, the response of the silver films can no longer be described properly by a sheet model. A \unit{25}{\nano\meter} silver film has an electric sheet conductivity of $\sigma_{//}^{(\mathrm{e})} = 0.48/\eta_0 = \unit{1262}{\micro\siemens}$, to be compared with the sheet conductivity of graphene equal to $\sigma_\mathrm{g} = \unit{60.8}{\micro\siemens}$. So even for relatively thin films, the current transport in silver films is more efficient than in graphene at infrared and optical frequencies.

\begin{figure}
\begin{center}
\includegraphics[clip]{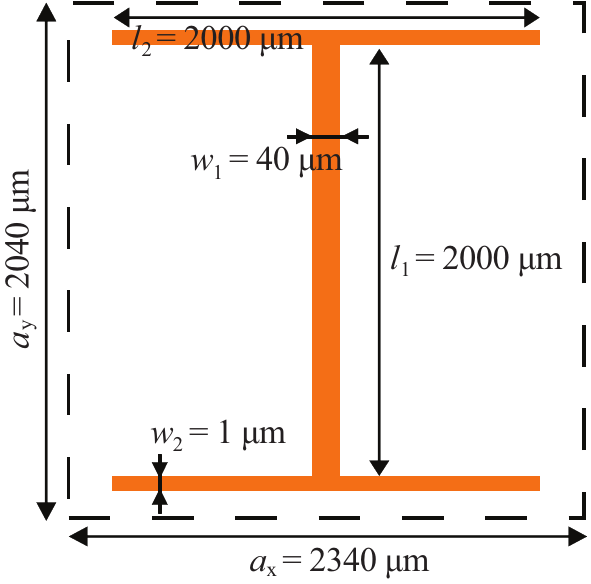}
\end{center}
\caption{Unit cell of the single-layer cut-wire metamaterial considered here; the dashed line indicates the unit cell boundaries. The wire has large end caps to keep the lattice constant sufficiently below the wavelength.}
\label{Fig:CutWire}
\end{figure}

\section{Application to a single-layer metamaterial}
\label{Sec:CutWires}
\begin{figure}
\begin{center}
\includegraphics[clip]{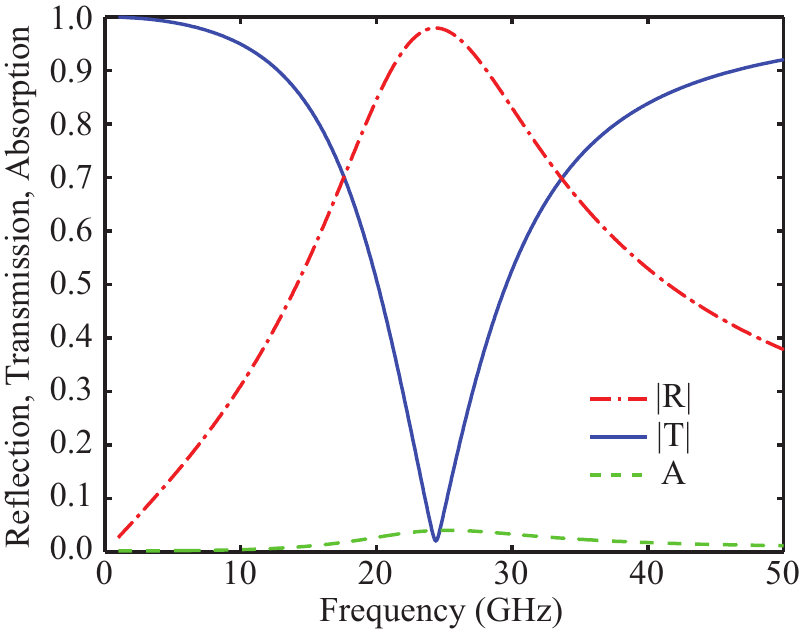}
\end{center}
\caption{Scattering parameters and absorption of the single-layer cut-wire metamaterial considered here. A resonance around \unit{24}{\giga\hertz} is observed.}
\label{Fig:ScatteringParamsCutWire}
\end{figure}

As a second example, we determine the effective sheet conductivity of a single-layer metamaterial. The unit cell of the metamaterial, shown in Fig.~\ref{Fig:CutWire}, consists of a simple cut-wire made from copper. The cut-wire has a quasistatic electric dipole resonance. We have calculated the scattering parameters using a frequency-domain electromagnetics solver (CST Microwave Studio). One unit cell is simulated with mirror symmetry boundary conditions (PEC/PMC) to model a periodic array of cut-wires. The resulting scattering parameters and the absorption of the metamaterial are plotted in Fig.~\ref{Fig:ScatteringParamsCutWire}. We see that there is a resonance---the electric dipole resonance---at \unit{25}{\giga\hertz} where the incident wave is highly scattered and we observe a peak in the absorption around the same frequency.

\begin{figure}
\begin{center}
\includegraphics[clip]{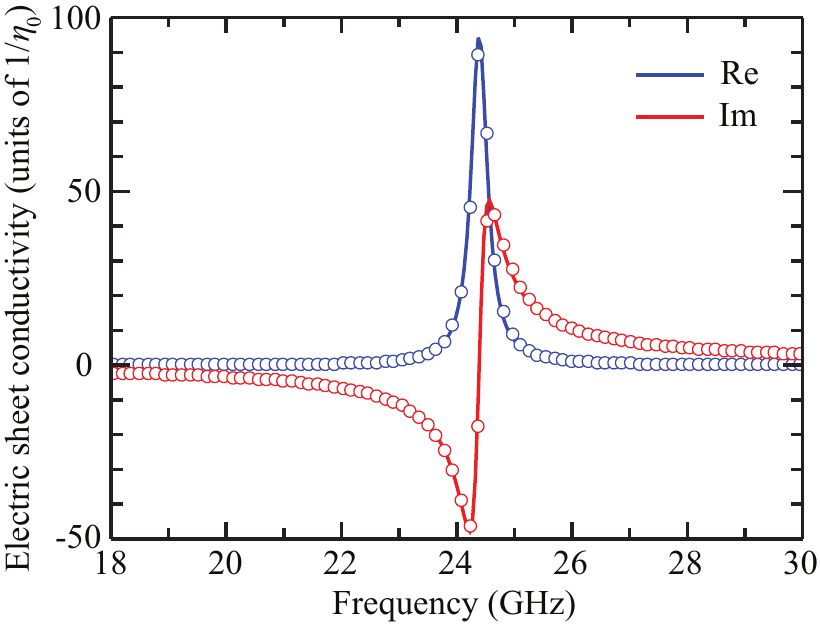}
\end{center}
\caption{Retrieved electrical conductivity of the single-layer cut-wire metamaterial. The full lines are the retrieved conductivity; the circles are a Lorentzian curve fitted onto the retrieved conductivity function.}
\label{Fig:ConductivityCutWire}
\end{figure}

The retrieved electric sheet conductivity following Eqs.~(\ref{Eq:RetrievalElectric})-(\ref{Eq:RetrievalMagnetic}) is plotted in Fig.~\ref{Fig:ConductivityCutWire}. (The magnetic sheet conductivity is several order of magnitudes smaller than the electric sheet conductivity.) The electric sheet conductivity has a clear Lorentzian resonance, a fact that we have checked by fitting a Lorentzian function,
\begin{equation}
\sigma_{//}^{(\mathrm{e})} = \frac{\mathrm{i}\kappa f}{f^2 + \mathrm{i}\Gamma f - f_\mathrm{r}^2} - \mathrm{i}\beta f,
\end{equation}
to the complex conductivity (the modification of the Lorentzian is because the current is the time derivative of the dielectric polarization). The Lorentzian spectrum fits the retrieved conductivity almost perfectly with resonance frequency $f_\mathrm{r} = \unit{24.3}{\giga\hertz}$, damping frequency $\Gamma = \unit{0.36}{\giga\hertz}$, coupling constant $\kappa = \unit{88.3}{\mega\hertz}$, and background polarization $\beta = \unit{1.89}{\femto\second}$. The excellent fit of this Lorentzian spectrum to the electric conductivity confirms the quasistatic nature of the electric dipole resonance. The latter fact, together with the negligible magnetic conductivity, confirms that a single-layer cut-wire metamaterial can be almost perfectly described by the sheet model.

\section{Conclusions}
\label{Sec:Conclusions}
In this paper, we have detailed a retrieval approach for thin-film systems by mapping the scattering parameters (reflection and transmission) onto the electric and magnetic sheet conductivities of a two-dimensional current sheet. We have applied the sheet retrieval method to two nanophotonic systems. For thin films, we find the sheet retrieval to be valid for films with thicknesses up to \unit{25}{\nano\meter} at $\lambda = \unit{1.55}{\micro\meter}$ and we find the sheet conductivity of these silver films to be larger than the conductivity of graphene. For the single-layer cut-wire metamaterial, we demonstrate that the effective sheet conductance is an excellent representation of the scattering response. We believe that the sheet retrieval may be more appropriate for many single-layer metamaterials, because it does not require any model parameter (e.g., the thickness) to be determined. 

\section*{Acknowledgements}
Work at Ames Laboratory was partially supported by the U.S.\ Department of Energy, Office of Basic Energy Science, Division of Materials Sciences and Engineering (Ames Laboratory is operated for the U.S.\ Department of Energy by Iowa State University under Contract No.\ DEAC02-07CH11358) and by the U.S.\ Office of Naval Research, Award No.\ N000141010925. Work at FORTH was supported by the European Community's FP7 projects NIMNIL, Grant Agreement No.\ 228637.

\end{document}